\documentclass{sigchi-ext}
\usepackage[T1]{fontenc}
\usepackage{textcomp}
\usepackage[scaled=.92]{helvet} 
\usepackage{graphicx} 
\usepackage{balance}  
\usepackage{booktabs} 
\usepackage{ccicons}  
\usepackage{ragged2e} 
\usepackage{pgfplots}
\usetikzlibrary{patterns}

\def\plaintitle{The Struggle is Real: Analyzing Ground Truth Data of TLS (Mis-)Configurations} 
\def\emptyauthor{}
\def\plainkeywords{TLS Configuration; Qualys; Usability; Security; }

\title{The Struggle is Real: Analyzing Ground Truth Data of TLS (Mis-)Configurations}

\numberofauthors{2}
\author{%
  \alignauthor{%
    \textbf{Christian Tiefenau}\\
    \affaddr{University of Bonn} \\
    \affaddr{Bonn, Germany} \\
    \email{tiefenau@cs.uni-bonn.de} 
  }
  \vfil 
  \alignauthor{%
    \textbf{Emanuel von Zezschwitz}\\
    \affaddr{University of Bonn}\\
    \affaddr{Bonn, Germany}\\
    \email{zezschwitz@cs.uni-bonn.de} 
  }
}

\definecolor{linkColor}{RGB}{6,125,233}
\hypersetup{%
  pdftitle={\plaintitle},
  pdfauthor={\emptyauthor},
  pdfkeywords={\plainkeywords},
  bookmarksnumbered,
  pdfstartview={FitH},
  colorlinks,
  citecolor=black,
  filecolor=black,
  linkcolor=black,
  urlcolor=linkColor,
  breaklinks=true,
}

\begin{document}

\copyrightinfo{Copyright is held by the author/owner. Permission to make digital or hard
copies of all or part of this work for personal or classroom use is granted
without fee. Poster presented at the 14th Symposium on Usable Privacy and
Security (SOUPS 2018).}

\maketitle

\RaggedRight{} 

\begin{abstract}
  As of today, TLS is the most commonly used protocol to protect communication content. To provide good security, it is of central importance, that administrators know how to configure their services correctly. For this purpose, services like, e.g., \textit{Qualys SSL Server Test} can be leveraged to test the correctness of a given web server configuration.
  We analyzed the utilization of this service over a period of 2.5 months and found two major usage-patterns. In addition, there is a relation between the number of test-runs and the resulting quality (i.e., security) of a TLS configuration. 
\end{abstract}

\keywords{\plainkeywords}
\category{H.5.m}{Information interfaces and presentation (e.g.,
  HCI)}{Miscellaneous}

\section{Introduction}
One of today's most relevant cryptographic protocols is Transport Layer Security (TLS). It provides communication security by symmetrically encrypting the data between two parties that have access to a shared secret which is generated at the beginning of a conversation. TLS is widely used in applications that provide secure communication over the internet. For example, web browsers, email, smartphone applications and the more recent class of smart home devices (e.g., Amazon Alexa) rely on the security of correct TLS configurations. 
However, recent studies on TLS \cite{durumeric2013analysis, holz2015tls} show, that many endpoints are vulnerable to publicly known attacks like ``Heartbleed''\footnote{\url{https://heartbleed.com}} or ``DROWN'' \cite{aviram2016drown}. In addition, especially web servers are often poorly configured and allow for Man-in-the-Middle attacks~\cite{fahl2014eve}.

\begin{marginfigure}[0pc]
  \begin{minipage}{\marginparwidth}
    \centering
    \begin{tikzpicture}
    
      \begin{axis}[
        height=9cm,width=\marginparwidth,
        xbar, enlargelimits=0.15, xmin=0, xmax=60,
        xlabel={Percentage},
        symbolic y coords={.br,.uk,.jp,.de,.net,.org,.ru,.com},
        ytick=data,
        nodes near coords, nodes near coords align={horizontal},
		legend columns=2,
		legend style={at={(0.5,-0.2)},anchor=north},
        ]
        \addplot[pattern=crosshatch dots]  coordinates {(43.3,.com) (3.6,.org) (2.1,.ru)  (4.7,.net) (4.6,.de) (2.8,.uk) (1.4,.br) (1.1,.jp)};
        \addplot[pattern=grid] coordinates {(46.4,.com) (5.1,.org) (4.7,.ru)  (4.0,.net) (3.6,.de) (2.2,.uk) (2.0,.br) (1.7,.jp)};
        \legend{Our Data, W3Techs}
      \end{axis}
    \end{tikzpicture}
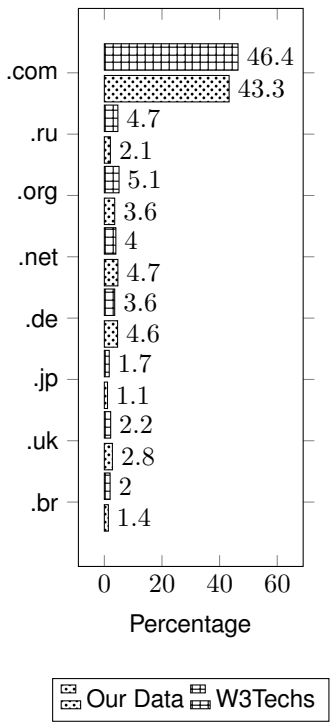
    \caption{The distribution of the most used top level domains in our sample}~\label{fig:tldpercentagew3}
  \end{minipage}
\end{marginfigure}

Fahl et al. \cite{fahl2014eve} found that webmasters often lack a detailed understanding of TLS which results in insecure and vulnerable communication channels. A recent lab experiment related to the configuration process of TLS (i.e., enabling HTTPS) \cite{krombholz2017have} confirmed, that even experienced users have problems to execute the configuration task in the most secure way. Krombholz et al. \cite{krombholz2017have} reported poor usability as the main reason for weak configuration results. While this is a important finding as it indicates that administrators, and thus the security of a TLS configuration, would benefit from more usable processes, the researchers excluded important real-world aspects. Most importantly, the evaluation was based on a short-term lab experiment and focused on the initial deployment process of a HTTPS-enabled web server configuration and thus, excluded long-term maintenance tasks, such as certificate renewal and the administrators' reactions to newly discovered vulnerabilities.  

In contrast to this previous work, we analyzed real TLS security test results provided by the web service \textit{Qualys}\footnote{\url{https://www.ssllabs.com/ssltest/}} and propose to use this data for a complementary research approach. \textit{Qualys} is a web service which assesses the vulnerability of a TLS server configuration through grades from A to F. We argue that analyzing such data can provide important insights into real-world problems concerning TLS configurations and can be used to confirm or question previous knowledge gained from lab studies and self-reports. 

Over a time span of 2.5 months, we collected 445,187 test results from 144,292 unique domains.
We identified two usage patterns: the \textit{Initial Configuration} and \textit{Reoccurring Checks}. Regarding the number of tests, we observed that the grades are increasing when more tests are done. In summary, this paper provides preliminary real-world insights into the difficult task of configuring TLS and indicates that more usable concepts are needed. In addition, we confirm previous findings which found that expert users are often unable to decide on the appropriate level of security. Our results highlight the need for more usable configuration processes and feasible support in decision making.

\section{Approach}
To collect real-world data, we used one of the most popular online service for SSL testing from \textit{Qualys SSL Labs}. We specifically focused on the \textit{Recently Seen}, the \textit{Recent best} and the \textit{Recent worst} grades which were shown on the services' front page. We wrote a python script that visited the site every five seconds\footnote{During the development phase, five seconds was found to be optimal for our purpose.}, extracted the domains with their related grades and stored them in a database. We collected data from September, 29th 2017 to December 14th, 2017, a time period of about 2.5 months (77 days). Overall, we analyzed 269,097 tests from 143,660 unique domains.

\subsection{Qualys Rating Scale}
\textit{Qualys} scoring is a mix of three categories which influence the security of a servers' TLS configuration. According to \textit{Qualys}, the final grade is calculated based on the following aspects:
\begin{itemize}
    \item \textit{Protocol support (30\%)}: Depending on the supported protocols (SSL 2.0, SSL 3.0, TLS 1.0, TLS 1.1, TLS 1.2) the grade is defined by the weakest one available.
    \item \textit{Key exchange (30\%)}: Based on the parameters at key exchange (e.g. key length).
    \item \textit{Cipher strength (40\%)}: Qualys scores the length of the used cipher.
\end{itemize}

\begin{margintable}[-8pc]
  \begin{minipage}{\marginparwidth}
    \centering
  \begin{tabular}{l r}
    \small\textit{Sessions} & \small \textit{Occurrences} \\
    \midrule
    1 & 132,396 \\
    2 & 7,399 \\
    3 & 1,950 \\
    4 & 826 \\
    5 & 386 \\
    $\geq$6 & 669  \\
    \midrule
    \textbf{Total} & \textbf{143,626} 
  \end{tabular}
  \caption{Number of clustered testing sessions per domain}~\label{tab:sessiondistribution}
  \end{minipage}
\end{margintable}

\begin{margintable}[0pc]
  \begin{minipage}{\marginparwidth}
    \centering
    \begin{tabular}{l r r}
        {\small\textit{Grade}} & {\small \textit{Total}} & {\small \textit{Percentage}}\\
        \midrule
        A+ (6) & 17,427 & 14.5\% \\
        A  (5) & 50,630 & 42.0\% \\
        A- (4) & 11,976 & 10.0\% \\
        B  (3) & 14,759 & 12.2\% \\
        C  (2) & 9,152 & 7.6\% \\
        F  (1) & 10,047 & 8.3\% \\
        T  (0) & 6,550 & 5.4\% \\
        \midrule
        \textbf{Total} & \textbf{120.541} & 
    \end{tabular}
    \caption{The distribution of the last grade we observed for each domain}~\label{tab:lastgradedist}
  \end{minipage}
\end{margintable}

The combination of these three categories results in a final score from 0 to 100, that gets translated in a grade from A+ to F.
More details about the used scoring rules (e.g., the differences of A+, A and A-) are available online\footnote{\url{https://github.com/ssllabs/research/wiki/SSL-Server-Rating-Guide} -- accessed: 05/24/2018}.

\section{Preliminary Results}
Even though our observation was limited to 77 days and thus represents a snap shot of the real world, we are confident that the data is suited to inform our research question. Fahl et al. \cite{fahl2014eve} observed that 13.6\% of all servers triggered a warning in the users' browser. Our data confirms this finding as 13.7\% of all servers were graded with F or T, representing a very weak configuration (e.g., use a broken cipher like RC4, a key length of 512 and SSL 2 enabled) and a non-trusted certificate. Such configurations usually trigger warning messages in modern web browsers. 
In addition, the Top-Level-Domain (TLD) distribution of our dataset matches the overall distribution as compared to publicly available W3Techs data \footnote{\url{https://w3techs.com/technologies/overview/top_level_domain/all} -- accessed: 05/24/2018} (see figure \ref{fig:tldpercentagew3}).

\subsection{Usage}
In 77 days, we observed 269,097 tests from 143,660 domains. The average number of tests per domain was $\mu=1.87$ ($\sigma=32.86, min=1, max=6,857$). Since a small number of domains was responsible for a large number of tests, we opted to sanitize the data by focusing on the domains whose test count was in the interval of $[0, \mu+3\sigma = 160]$. This way, 34 domains were identified as outliers and excluded from the analyses. The final data set comprised 143,626 domains and 219,877 tests ($\mu=1.53, \sigma=2.53$ tests per domain).

We grouped multiple consecutive tests into the same \emph{Test Session}. As previous lab experiments reported configuration sessions of four hours \cite{krombholz2017have}, a session was defined as closed whenever we did not observe a test within four hours after the last test had been logged. Table \ref{tab:sessiondistribution} shows the number of sessions we observed and their corresponding occurrences. It shows that the vast majority of users performed just one test session within the given time frame of 77 days.

\textbf{Initial Configuration}\\
\textit{Initial Configurations} are defined as the first occurrence of a domain in the data set having at least two checks. We set a minimum of two checks as we were interested in evaluating iterative configuration processes. We assume that this data represents the process where an administrator tests the server multiple times until she is satisfied with the resulting grade and the properties of the configuration.

The vast majority - 132,396 (92.18\%) - just tested once in their first appearing four-hour-time-interval, indicating no iterative process. In the remaining 11,225 \textit{initial phases}, we observed 2.97 ($\sigma = 1.72$) tests on average, meaning that most people checked their server configuration three times before seeing their task finished or giving up. The maximum of performed checks within a test session was 25.

\textbf{Reoccurring Checks}\\
In addition, we found that some administrators use the Qualys service more frequently over longer periods of time. For example, one domain checked its configuration approximately every 2-5 minutes resulting in 6,857 checks within 2.5 months (\textasciitilde90 tests/day). Table \ref{tab:sessiondistribution} gives more details.

\begin{marginfigure}[0pc]
  \begin{minipage}{\marginparwidth}
    \centering
          \begin{tikzpicture}
            \begin{axis}[
                width=1.15*\marginparwidth, height=10cm,
                stack plots=y,
                area style,
        		enlarge x limits=false,
                transpose legend,
		        legend columns=2,
		        legend style={at={(0.5,-0.2)},anchor=north},
                xlabel={Number of tests},
                symbolic x coords={2, 3, 4, 5, 6, 7, 8, 9, $\geq$10},
                xtick=data,
                ]
            \addplot[pattern=horizontal lines, pattern color=green] coordinates {(2,12.95) (3,16.98) (4,17.62) (5,23.71) (6,24.51) (7,27.51) (8,25.11) (9,25.60) ($\geq$10,25.55)} \closedcycle;
            \addplot[pattern=vertical lines, pattern color=green] coordinates {(2,40.22) (3,39.53) (4,39.51) (5,39.07) (6,39.77) (7,37.23) (8,40.06) (9,37.68) ($\geq$10,43.77)} \closedcycle;
            \addplot[pattern=grid, pattern color=green] coordinates {(2,10.87) (3,10.22) (4,10.96) (5,8.74) (6,9.71) (7,9.60) (8,9.45) (9,10.39) ($\geq$10,9.63)} \closedcycle;
            \addplot[pattern=crosshatch, pattern color=yellow] coordinates {(2,12.63) (3,10.95) (4,11.28) (5,11.38) (6,9.93) (7,10.87) (8,10.44) (9,12.32) ($\geq$10,8.86)} \closedcycle;
            \addplot[pattern=north east lines, pattern color=orange] coordinates {(2,8.39) (3,6.50) (4,7.51) (5,5.54) (6,6.66) (7,4.51) (8,5.78) (9,6.04) ($\geq$10,5.20)} \closedcycle;
            \addplot[pattern=dots, pattern color=red] coordinates {(2,9.07) (3,9.10) (4,9.26) (5,6.66) (6,5.93) (7,5.09) (8,6.49) (9,4.83) ($\geq$10,3.98)} \closedcycle;
            \addplot[pattern=fivepointed stars, pattern color=red] coordinates {(2,5.87) (3,6.72) (4,3.87) (5,4.89) (6,3.49) (7,5.20) (8,2.68) (9,3.14) ($\geq$10,3.02)} \closedcycle;
            \legend{A+,A,A-,B,C,F,T}
            \end{axis}
        \end{tikzpicture}
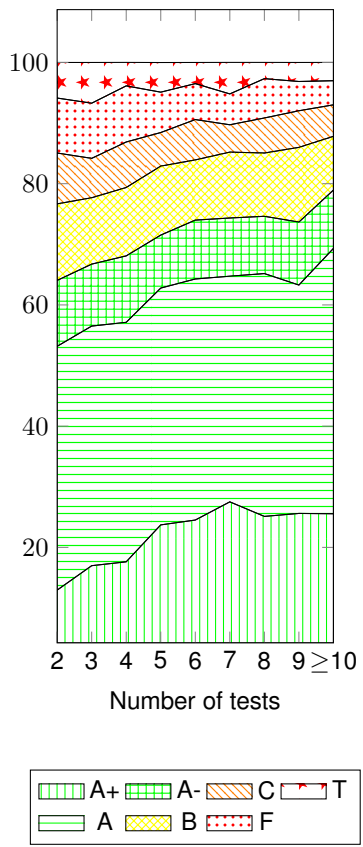
    \caption{Grade Distribution compared to the number of checks done with Qualys SSL Test}
    \label{fig:gradecountdistribution}
  \end{minipage}
\end{marginfigure}

\subsection{Grade Distribution}
To get insights into the overall distribution of the final grades, we focused on the last observed grade domains that were successfully tested at least twice\footnote{Qualys' test sometimes fails (e.g. due to ``Certificate name mismatch'') and requires a manual restart of the test process.}. 
Overall, we logged 120.541 successfully performed tests. Table \ref{tab:lastgradedist} shows the distribution of the grades. The average of all ``final'' ratings is close to A- (3.97) with a $\sigma$ of 1.75. 

\subsection{Grade changes}
Figure \ref{fig:gradecountdistribution} shows that with an increasing number of tests the proportion of all A grades rises from 64\% concerning two tests to nearly 75\% with six tests. This indicates that administrators tend to test and reconfigure their servers multiple times to achieve good security and tools like Qualys can support them in doing so.

\section{Discussion}
In the following, we discuss the implications of our preliminary findings.

\subsection{Securing TLS Configurations Takes Time}
Hardening TLS Configurations is an iterative process. Our analysis shows that the resulting TLS grade and thus, the servers' security is improving with the number of tests. This indicates that short-term lab studies might oversee some aspects when testing the ability of administrators to configure and harden a service that uses TLS as administrators often follow a trial and error approach and secure configurations may take several hours. 

\subsection{Not All Services Achieve Best Grades}
Despite the fact that Qualys suggests improvements that can be done to harden the security of a configuration, we found that more than 21\% of all logged domains finished their last test with a grade of C or worse. To get more insights into the reasons for this real-world behavior, future studies should collect qualitative feedback to analyze why some administrators seem satisfied with bad grades. 

\subsection{Only a Small Subset of Webservices Make Use of Qualys}
The number of all publicly known active websites\footnote{\url{https://www.netcraft.com/active-sites/} -- accessed: 05/24/2018} in combination with \textit{Let's Encrypt}'s statistics\footnote{\url{https://letsencrypt.org/stats/} -- accessed: 05/24/2018} indicates that there are nearly 115 million domains using TLS. We observed tests by only 0.12\% of these websites. Even though we assume that the number of administrators who use services like Qualys is much higher than observed, we still conclude that a minority of HTTPS-enabled web servers is actively tested and that the security of such services is seldom monitored. 

\section{Conclusion \& Future Work}
In this paper we proposed to analyze \textit{Qualys SSL Server Test} results as a new information vector to understand the problems of the TLS configuration process. 
In the future we plan to combine our quantitative data with qualitative feedback by contacting some of the observed web services directly.  
Finally, it would be interesting to observe and analyze testing behaviour whenever new TLS vulnerabilities like DROWN are found and publicly discussed.

\section{Acknowledgments}
The authors would like to thank Qualys Inc. for granting access to their services.

\balance{} 

\bibliographystyle{SIGCHI-Reference-Format}
\bibliography{main_ccs}


\begin{thebibliography}{00}


\ifx \showCODEN    \undefined \def \showCODEN     #1{\unskip}     \fi
\ifx \showDOI      \undefined \def \showDOI       #1{{\tt DOI:}\penalty0{#1}\ }
  \fi
\ifx \showISBNx    \undefined \def \showISBNx     #1{\unskip}     \fi
\ifx \showISBNxiii \undefined \def \showISBNxiii  #1{\unskip}     \fi
\ifx \showISSN     \undefined \def \showISSN      #1{\unskip}     \fi
\ifx \showLCCN     \undefined \def \showLCCN      #1{\unskip}     \fi
\ifx \shownote     \undefined \def \shownote      #1{#1}          \fi
\ifx \showarticletitle \undefined \def \showarticletitle #1{#1}   \fi
\ifx \showURL      \undefined \def \showURL       #1{#1}          \fi

\bibitem{aviram2016drown}
{Nimrod Aviram}, {Sebastian Schinzel}, {Juraj Somorovsky}, {Nadia Heninger},
  {Maik Dankel}, {Jens Steube}, {Luke Valenta}, {David Adrian}, {J~Alex
  Halderman}, {Viktor Dukhovni}, {and} {others}. 2016.
\newblock \showarticletitle{DROWN: Breaking TLS Using SSLv2}. In {\em USENIX
  Security}. 689--706.
\newblock


\bibitem{durumeric2013analysis}
{Zakir Durumeric}, {James Kasten}, {Michael Bailey}, {and} {J~Alex Halderman}.
  2013.
\newblock \showarticletitle{Analysis of the HTTPS certificate ecosystem}. In
  {\em IMC'13}. ACM, 291--304.
\newblock


\bibitem{fahl2014eve}
{Sascha Fahl}, {Yasemin Acar}, {Henning Perl}, {and} {Matthew Smith}. 2014.
\newblock \showarticletitle{Why eve and mallory (also) love webmasters: A study
  on the root causes of SSL misconfigurations}. In {\em CCS'14}. ACM, 507--512.
\newblock


\bibitem{holz2015tls}
{Ralph Holz}, {Johanna Amann}, {Olivier Mehani}, {Matthias Wachs}, {and}
  {Mohamed~Ali Kaafar}. 2015.
\newblock \showarticletitle{TLS in the wild: an Internet-wide analysis of
  TLS-based protocols for electronic communication}.
\newblock {\em arXiv:1511.00341\/} (2015).
\newblock


\bibitem{krombholz2017have}
{Katharina Krombholz}, {Wilfried Mayer}, {Martin Schmiedecker}, {and} {Edgar
  Weippl}. 2017.
\newblock \showarticletitle{``I Have No Idea What I'm Doing''-On the Usability
  of Deploying HTTPS}. In {\em USENIX Security}, Vol.~17. 1339--1356.
\newblock


\end{thebibliography}

\end{document}